\documentclass[12pt,preprint]{aastex}
\shorttitle{Stellar Rotation in Young Clusters I.}
\shortauthors{Huang \& Gies}

\begin{document}

\received{}
\accepted{}

\title{Stellar Rotation in Young Clusters. \\ 
I. Evolution of Projected Rotational Velocity Distributions}

\author{W. Huang and D. R. Gies\altaffilmark{1}}

\affil{Center for High Angular Resolution Astronomy\\
Department of Physics and Astronomy \\
Georgia State University, P. O. Box 4106, Atlanta, GA  30302-4106;\\
huang@chara.gsu.edu, gies@chara.gsu.edu}

\altaffiltext{1}{Visiting Astronomer, Kitt Peak National Observatory
and Cerro Tololo Interamerican Observatory, 
National Optical Astronomy Observatory, operated by the Association
of Universities for Research in Astronomy, Inc., under contract with
the National Science Foundation.}

\slugcomment{Submitted to ApJ}

\paperid{63886}

%%%%%%%%%%%%%%%%%%%%%%%%%%%%%%%%%%%%%%%%%%%%%%%%%%%%%%%%%%%%%%%

\begin{abstract}
Open clusters offer us the means to study stellar properties 
in samples with well-defined ages and initial chemical composition. 
Here we present a survey of projected rotational velocities
for a large sample of mainly B-type stars in young 
clusters to study the time evolution of the rotational properties of 
massive stars.  The survey is based upon moderate resolution spectra 
made with the WIYN 3.5~m and CTIO 4~m telescopes and Hydra multi-object 
spectrographs, and the target stars are members of 
19 young open clusters with an age range of approximately 6 to 73~Myr.
We made fits of the observed lines \ion{He}{1} 
$\lambda\lambda 4026, 4387, 4471$ and \ion{Mg}{2} $\lambda 4481$ 
using model theoretical profiles to find projected rotational velocities 
for a total of 496 OB stars.  We find that there are fewer slow rotators
among the cluster B-type stars relative to nearby B stars in the field. 
We present evidence consistent with the idea that the more massive B stars 
($M > 9 M_\odot$) spin down during their main sequence phase.  
However, we also find that the rotational velocity 
distribution appears to show an increase in the numbers of rapid rotators
among clusters with ages of 10~Myr and higher.  These rapid rotators 
appear to be distributed between the zero age and terminal age main sequence
locations in the Hertzsprung-Russell diagram, and thus only a 
minority of them can be explained as the result of a spin up at 
the terminal age main sequence due to core contraction.  
We suggest instead that some of these rapid rotators may have been 
spun up through mass transfer in close binary systems.  
\end{abstract}


%%%%%%%%%%%%%%%%%%%%%%%%%%%%%%%%%%%%%%%%%%%%%%%%%%%%%%%%%%%%%%%

\setcounter{footnote}{1}
\section{Introduction}                              % Section 1

Unlike low mass stars that can lose their angular momentum early  
in their main sequence phase, massive OB stars are usually born with 
a large initial angular momentum that can last
throughout their relatively short, core hydrogen-burning phase. 
The spin evolution of a single, non-magnetic, massive star is driven by  
angular momentum loss in the stellar wind, a net increase in the moment 
of inertia, and an increase in stellar radius.  Theoretical models of these
processes were recently developed independently by \citet{heg00}
and \citet{mey00}.  These models predict the changes
in the stellar interior, the equatorial rotational
velocity, and photospherical chemical abundances 
that occur along the evolutionary tracks.
One of the predictions of these models is that  massive stars spin down during
core H-burning at a rate that is larger for more massive stars and for those with 
higher initial rotational speeds.  These models predict that a spin-up episode can 
occur very close to the terminal age main sequence (TAMS)  
when the core contracts prior to H-shell burning. In some situations 
this increase may lead to near critical rotational velocities
and induce enhanced equatorial mass outflows such as are observed in rapidly 
rotating Be stars \citep*{mae99,zor04}.  \citet{fab00} 
found that the incidence of Be stars tends to peak in open clusters with ages 
between 13 and 25 Myr, approximately when early B-type stars are close to the TAMS.
A similar conclusion was also found by \citet{mcs05} based upon a study 
of the frequency of Be stars in some 48 southern sky clusters.  
However, the positions of the Be stars in the Hertzsprung-Russell diagram (HRD)
cover the entire range between the zero-age main sequence (ZAMS) and the TAMS
\citep{mer82,zor04,mcs05} so it is doubtful that many Be stars are actually 
TAMS objects. 

Another possible process causing spin-up in massive stars is mass transfer in 
close binary systems.  There are many examples of rapid rotation of the mass 
gainer in close binaries \citep{vdh70,etz93,sim99,bar04}. 
Even moderate mass transfer can cause a significant 
spin up of the mass gainer star \citep{lan04}.  One extreme example of this spin up
process is found in the classical Be star, $\phi$ Persei, which was probably spun
up by an earlier stage of mass transfer to attain the observed projected 
rotational velocity of $V \sin i$ = 450 km~s$^{-1}$ (about 97\% of the critical velocity; 
\citealt{gie98}). 

Young star clusters provide us with examples of stellar populations 
that probably started with the same chemical composition and are now observed 
with various well-defined ages.  Spectroscopic analysis of the stars in young clusters 
offers us the means to examine these various processes affecting rotation. 
Here we present an investigation of the stellar rotation properties of 
the brighter stars in 19 open clusters with an age range between 6 and 73~Myr. 
In \S2 we describe the observing runs and data reduction procedures, and
in \S3 we outline the details of the line synthesis method we used to 
determine the stellar projected rotational velocity $V \sin i$. 
We report in \S4 on the statistical results based on
the measured $V \sin i$ and some implications of these results
for evolutionary models. 

%%%%%%%%%%%%%%%%%%%%%%%%%%%%%%%%%%%%%%%%%%%%%%%%%%%%%%%%%%%%%%%

\section{Observations}                                   % Section 2

We obtained moderate dispersion, blue spectra of B-type stars 
in 19 Galactic clusters.  We used the WEBDA open cluster 
database\footnote{http://www.univie.ac.at/webda/}
\citep{mer03} to select clusters for this study based on the following
criteria: (1) age $<75$~Myr, (2) distance modulus $m-M< 15$,
(3) angular diameter 10 -- 60 arcmin, and 
(4) more than 20 main sequence and/or giant star members with
MK classifications of B9 or earlier.  
The names of these clusters are listed below (Table~4). 
The spectra of the northern sky clusters were obtained with the
WIYN 3.5~m telescope at Kitt Peak National Observatory while
the southern clusters were observed with the Cerro Tololo
Interamerican Observatory 4~m telescope.  Both telescopes are
equipped with Hydra multiple-object spectrographs \citep{bar95}
that use micropositioners and fiber optics to obtain spectra
of many targets simultaneously.  Our main goal was to record the
absorption line profiles of \ion{He}{1} $\lambda\lambda 4026, 4387, 4471$, 
\ion{Mg}{2} $\lambda 4481$, and H$\gamma$ $\lambda 4340$ that are 
relatively strong throughout the B-type spectral class.  
We generally observed each cluster field on at least two nights,
so that we could make a preliminary identification of the radial
velocity variable stars (candidate spectroscopic binaries) 
and double-lined binaries (whose line profiles may mimic 
those of rapid rotators at some orbital phases).  

The details of the spectrograph set up for each run are 
summarized in Table~1.  The Hydra spectra from the two runs 
are comparable but the WIYN spectra have better resolving power 
than those from CTIO.  All the targets in a particular cluster 
were observed in a single fiber configuration using multiple 
exposures of 10 minutes (CTIO) or 20 minutes (WIYN) duration. 
Additional fibers were set on the blank sky to estimate 
the sky background.  In addition to nightly bias and dark
frames, we also obtained flat field spectra for every fiber 
configuration and telescope pointing.  We typically obtained 
spectra of 30 -- 80 stars in each observation, but we report 
here only on the brighter B-type stars that we identified in 
each field. 
 
\placetable{tab1}      % Table 1 - Observing runs 

All the spectra and calibration frames were reduced using 
standard routines in IRAF\footnote{IRAF is distributed by the
National Optical Astronomy Observatory, which is operated by
the Association of Universities for Research in Astronomy, Inc.,
under cooperative agreement with the National Science Foundation.}.
The extraction and calibration of the individual spectra 
from the images was done using the {\it dohydra} package 
\citep{val92}, in which flat fielding is accomplished by 
dividing the stellar spectrum by the flat field spectrum 
after extraction.  The individual sub-exposures were extracted
separately and then co-added to improve the S/N ratio.  
All the spectra were transformed to a uniform wavelength
scale and rectified to a unit continuum, and most have a 
S/N = 50 -- 400 pixel$^{-1}$ (depending on the star's brightness). 
The focus of each spectrograph varied slightly across the spectrum, 
and we used width measurements of the unresolved comparison 
lines to estimate the FWHM of the resolution element (yielding 
the instrumental spectral line broadening) at the positions 
of the important spectral features.  These are listed in the
last three rows of Table~1.  

%%%%%%%%%%%%%%%%%%%%%%%%%%%%%%%%%%%%%%%%%%%%%%%%%%%%%%%%%%%%%%%

\section{Measuring Projected Rotational Velocities}       % Section 3

Rotational broadening is the process that dominates the apparent shapes of 
photospheric absorption lines in unevolved B-type stars, and the derivation of 
a projected rotational velocity ($V \sin i$) from their   
shape is the first step to study stellar rotation.  One simple
way to characterize these broadened line profiles is by measuring a line width 
(such as FWHM). The $V \sin i$ value can be directly determined if the relation between 
the line width and $V \sin i$ is carefully calibrated \citep{sle75,how97}. 
This method has been developed and improved over many decades, and it has 
become the dominant method in this field because of its simplicity and efficiency
for processing large amounts of spectral data.  However, since this method is based 
only upon the line width to gauge the effects of rotation on the line profile,
the accuracy of the results may suffer.  Furthermore, in the case of fast stellar rotation 
line width measurements become less useful due to the reduction in wing strength
caused by strong gravity darkening of the equatorial flux in the stellar atmosphere
\citep*{tow04}.  

One way to solve this problem and to obtain more reliable and accurate
$V \sin i$ measurements is by fitting the entire observed line profile 
with a grid of the theoretical calculations \citep{how01,mat02,tow04}. 
Because this method uses all the available line data, we can derive
a secure value for $V \sin i$ and help establish other 
physical properties such as effective temperature, gravity, and chemical 
abundance.  The key to the success of this method is to begin 
by constructing reliable theoretical line profiles. 
It is often assumed for the sake of simplicity that the 
local intensity profiles have a Gaussian or Lorentzian shape, 
and then flux profiles are calculated by integrating these shapes 
over the visible, limb-darkened stellar surface. 
However, in this paper we will mainly examine \ion{He}{1} lines that have specific intensity 
profiles that are too complicated to be described by Gaussian or Lorentzian functions.
Therefore, we decided to calculate realistic intensity profiles using 
the stellar atmosphere code TLUSTY and the radiative transfer code SYNSPEC 
\citep{hub95}. 

For hotter stars ($T_{\rm eff} >$ 13000~K), we choose to fit the \ion{He}{1} lines 
with the synthesized profiles based on intensity profiles calculated by TLUSTY and SYNSPEC,
while we modeled and fit the \ion{Mg}{2} $\lambda 4481$ line for the cooler, late-type B stars. 
We first calculated a set of simple H-He LTE model atmospheres using TLUSTY, 
assuming solar abundances and a constant microturbulent velocity (2 km~s$^{-1}$).
We then constructed a grid of specific intensity profiles using SYNSPEC that 
was based on a three dimensional parameterization using 
$\mu$, the cosine of the angle between the surface normal and line of sight, 
$T_{\rm eff}$, the local effective temperature, and 
$\log g$, the logarithm of the surface gravity. 
The effective temperature ranges from $T_{\rm eff}=12000$~K to 34000~K in steps of
2000$^\circ$ for the \ion{He}{1} lines (and from 8000~K to 20000~K for the \ion{Mg}{2} line); 
the surface gravity spans the range from $\log g = 3.2$ to 4.2 in steps of 0.2 dex; 
and the orientation cosine ranges from $\mu=0.1$ to 1.0 in steps of 0.1. 

We then integrated the flux contributions over the visible hemisphere of the 
model star to obtain the synthesized line profiles.  The synthesis procedure included: 
(1) dividing the stellar surface into a large number of facets with approximately 
equal area (10000 - 20000 facets);  
(2) finding $T_{\rm eff}$ and $\log g$ at the center of each facet, by 
considering the Roche geometry for a point-like (condensed) mass distribution and 
the consequent gravity darkening caused by stellar rotation \citep{col66};  
(3) calculating the specific intensity as a function of wavelength across 
the profile for each facet by interpolating within the grid of
specific intensity profiles described above; 
(4) integrating the contribution from all visible facets, Doppler-shifted
properly and weighted by the projected area; and
(5) rectifying the final output to a unit continuum.    

Note that these profiles are based upon simple model 
atmospheres that neglect line blanketing, so that our 
assigned effective temperatures may be somewhat hotter than the actual values.  
However, the change in the width of the intensity profiles caused 
by the adoption of an elevated temperature has a negligible impact on 
our derived projected rotational velocities because the rotational broadening 
is so much larger than the thermal and pressure broadening.  
We checked the reliability of this approach by calculating synthetic
model profile fits using both unblanketed and blanketed model 
atmospheres (from the Kurucz code ATLAS9 for 
Linux\footnote{http://www.hs.uni-hamburg.de/DE/Ins/Per/Reiners/package.htm}) 
for a sample of three stars that cover the observed range in 
projected line broadening, and we found that the derived values
of $V\sin i$ were identical within errors in each case. 
Thus, our use of simple LTE models is satisfactory for  
the purpose of determining $V \sin i$ from line profiles.   

We compared our model \ion{He}{1} and \ion{Mg}{2} profiles directly with 
those in the observed spectra to estimate the appropriate temperature model 
that we parameterized by a spectral subtype index.  Table~2 lists 
the physical parameters (effective temperature, mass, radius) for the model grid that were 
selected and/or interpolated to correspond approximately to a pseudo-spectral subtype 
according to the spectral subtype -- temperature calibrations given 
by \citet*{han97} (for the hotter subtypes) and by \citet{gra92} 
(for the cooler subtypes).   The \ion{He}{1} lines increase in 
strength to a maximum near subtype B2 and then begin to decline, 
but we can distinguish between the hot and cool portions of the 
\ion{He}{1} curve by the appearance of \ion{Mg}{2}, which declines
monotonically with increasing temperature in the B-star range.   We found the best fit 
profile from the grid for each available line, and we adopted the 
average spectral subtype index for models used in the rotational broadening
analysis.  We caution that these indices may not correspond to an 
actual MK subtype for a number of reasons:    
(1) we are assuming for fitting purposes that the stars are 
main sequence objects, but many are in fact evolved stars (i.e., giants);  
(2) the assumption of a He solar abundance will not be accurate 
for the He-peculiar B-stars;  
(3) there may be line blending problems for the spectroscopic binary targets; and 
(4) the theoretical line profiles are based on simple H-He LTE, 
unblanketed atmosphere models that may yield a slightly different 
spectral subtype -- temperature relation than adopted elsewhere. 
For example, we compared a series of synthetic profiles of \ion{He}{1} $\lambda 4471$
calculated from both unblanketed and line-blanketed main sequence models, 
and these show that a given line equivalent width is matched by 
an unblanketed model with a temperature that is $\approx 15\%$ higher 
than that for line-blanketed model atmospheres for B-stars with 
$T_{\rm eff} < 22000$~K (and about $\approx 4\%$ higher for stars 
hotter than this).  Therefore, the spectral subtype index from Table~2 
that is assigned to a target star may be of a slightly earlier type 
than would be determined by actual spectral classification.  
In the next paper in this series \citep{hua06}, 
we will fit the H$\gamma$ and \ion{He}{1} spectral lines using fully line blanketed
atmospheres in order to derive the temperature, gravity, and He abundance 
of each target.    
 
\placetable{tab2}      % Table 2 - Model ZAMS parameters as a function of subtype

We constructed rotational line profiles for each line and for each spectral 
subtype index over a grid of projected rotational velocities 
($V \sin i$ = 0, 50, 100, 150, 200, 250, 300, 350, and 400 km~s$^{-1}$) 
with a fixed inclination angle of $i=90^\circ$ 
(corresponding to a line of sight in the equatorial plane of the star). 
Finally, all the synthesized profiles were convolved 
with proper instrumental broadening profiles (Table 1) 
before comparison with the observations. 
For each spectrum we first estimated the spectral subtype index, 
determined an approximate projected rotational velocity, and 
then made a final fit after applying small corrections for the 
radial velocity shift and continuum placement.  
We found solutions for $V \sin i$ that minimized the $\chi^2$ difference 
between the observed and calculated profiles for the adopted spectral subtype, 
and then formed an average based on all the available lines. 
Examples of the fitting procedure are illustrated in Figure~1 
for a fast rotator ({\it top panels}) and a slow rotator ({\it bottom panels}) 
in the cluster NGC~884 ($\chi$ Per). 

\begin{figure}
%\epsscale{0.75}
%\plotone{f1.eps}
\plotfiddle{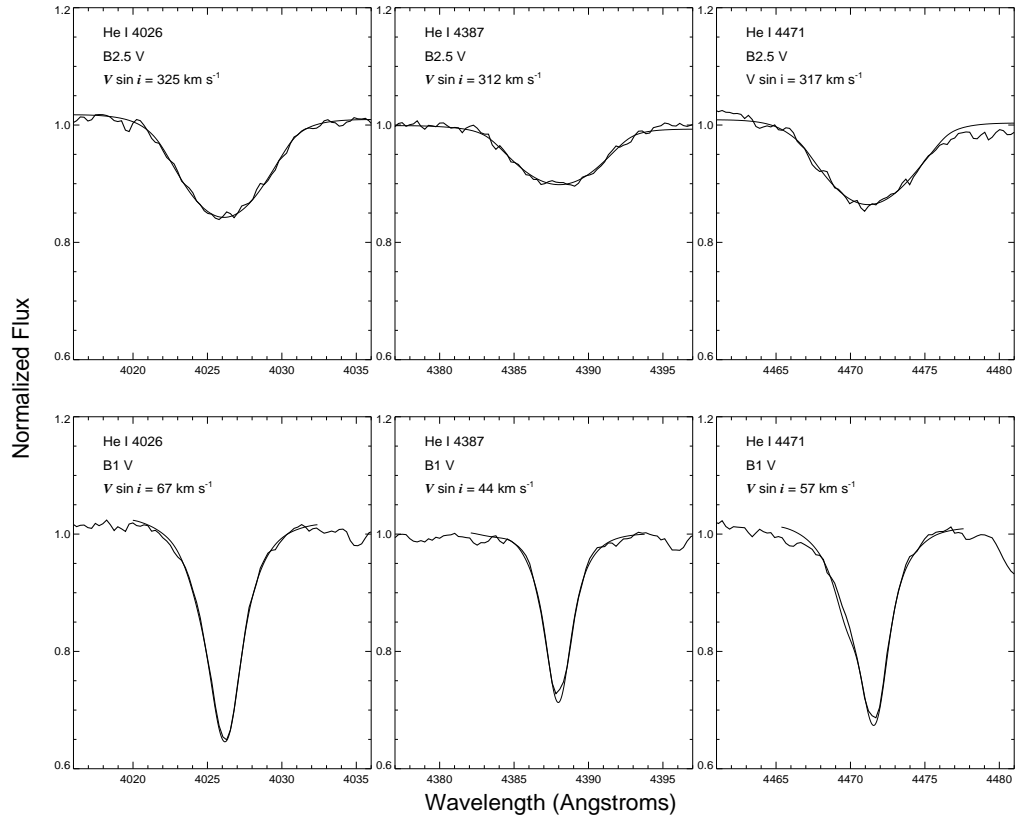}{100pt}{90}{360}{450}{30}{0}
\caption{
Fitting results for the \ion{He}{1} lines ({\it thin lines})
compared with observed profiles ({\it thick lines}).   
The top panels show fits for a fast rotator, NGC~884 \#2255, while 
the bottom panel shows the same for a slow rotator, NGC~884 \#1983.
}
\label{fig1}
\end{figure}

\placefigure{fig1}     % Figure 1 - profile fitting examples 

The measurement errors varied among our sample stars, depending on their magnitudes,
spectral subtypes, and observing set up.  Measurements of the \ion{He}{1} lines in 
spectra of stars of subtypes later than B7 are difficult because the 
lines are weaker and often blended with nearby metallic transitions. 
For the cooler B stars ($T_{\rm eff} \leq 13000$~K) we relied on fits of the 
\ion{Mg}{2} $\lambda 4481$ spectral region (which included 
\ion{Ti}{2} $\lambda 4468$, \ion{He}{1} $\lambda 4471$, 
\ion{Fe}{2} $\lambda 4473$, and \ion{Fe}{1} $\lambda 4476$).  
A comparison of the derived $V\sin i$ values among a sub-sample of 
late B-type stars where the \ion{He}{1} and \ion{Mg}{2} lines have 
comparable strength shows that we obtain the same projected 
rotational velocity within errors from fits of both line species. 
The average inter-line error on $V \sin i$ is about $10\%$ for the CTIO data and 
about $5\%$ for the WIYN data.  The systematic errors introduced by 
our choice of spectral subtype index are similar in size.  
The difference in the derived $V \sin i$ measurement is generally $<5\%$ when 
we compare results obtained using two adjacent spectral subtypes (see Table 2) 
and in fact decreases to zero around the B2 stars where the strength of
the \ion{He}{1} lines reaches a maximum and becomes insensitive to temperature.
The measured errors $\delta V \sin i$ for individual stars (Table~3, column 4) 
are estimated using the \ion{He}{1} $\lambda 4471$ or \ion{Mg}{2} $\lambda 4481 $ line fit. 
We define $\delta V \sin i$ as the offset from the best fit $V \sin i$ 
that increases the mean square of the deviations from the fit ${\rm rms}^2$ by
$2.7 {\rm rms}^2_{\rm min}/N$ (90\% confidence level), where $N$ is the number of
wavelength points within the fit region and ${\rm rms}^2_{\rm min}$ is 
the best fit minimum mean square of the deviations. 
Note that the resolution of our spectra (Table~1) limits our ability to 
measure $V\sin i$ accurately in very narrow-lined stars, so 
stars with $V\sin i < {\rm FWHM}/2$ probably have $V\sin i $ errors of 
approximately $\pm {\rm FWHM}/2$ (29 and 48 km~s$^{-1}$ for the WIYN and CTIO 
spectra, respectively). 

We found $V \sin i$ measurements in the WEBDA database 
on open clusters \citep{mer03} for 69 stars in common with our sample 
\citep{sle65,sle68,sle85,hil67,bal75a,dwo75,wol81,arn88,ver91,pen96,mat02}.
Some $90\%$ of these published $V \sin i$ values 
were obtained by simple visual estimates or by measuring line widths, but   
\citet{mat02} measured $V\sin i$ in 6 stellar spectra using line profile fitting 
and a simple convolution scheme for rotational broadening. 
Our $V \sin i$ values are generally in good agreement with these results (Fig.~2), 
which can be fit by a linear relation, 
\begin{equation} 
V \sin i ~[{\rm Published}] = (0.91\pm 0.05) ~V \sin i ~[{\rm Ours}]+(26\pm 8) ~{\rm km~s}^{-1}.
\end{equation} 
A large scatter is not unexpected in such a comparison because 
of the heterogeneity of the $V \sin i$ data sources and measurement techniques.
Our results are systematically larger than the published ones by about $9\%$, which
is probably due to our treatment of gravity darkening effects for fast rotators
that accounts for the weaker line wings.  The larger $V\sin i$ measurements for 
the published results in the low $V \sin i$ regime probably reflect the 
lower resolving power and S/N of the observations used in earlier studies. 

\begin{figure}
\epsscale{0.9}
\plotone{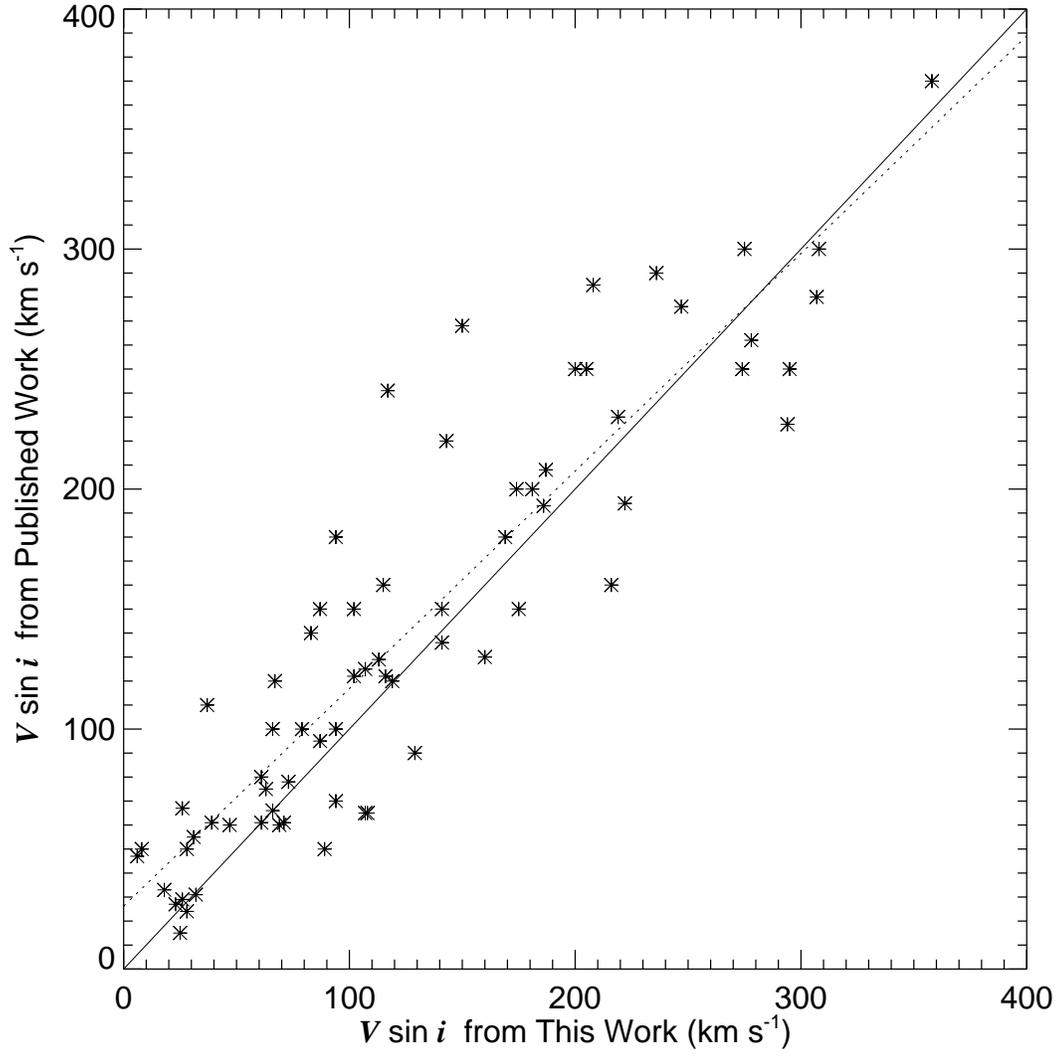}
\caption{
A comparison of our measured $V \sin i$ values with published work. 
The solid line shows the one-to-one relationship while the
dotted line is the result of a linear least-squares fit.
}
\label{fig2}
\end{figure}

\placefigure{fig2}     % Figure 2 - Comparison of V sin i measurements

We offer some cautionary remarks before consideration of our results in 
the next section.  The run of stellar parameters as a function of temperature or 
spectral subtype in Table 2 is probably reliable for slow rotators.
However, for fast rotators the physical conditions on the
stellar surface can be quite different from non-rotating models due to the rotational effects. 
For example, \citet{mey97} point out that the polar temperature of a fast rotator 
may be much higher than its effective temperature as derived from either the integrated 
luminosity and surface area or from the integrated spectrum of its visible hemisphere. 
Our scheme relies on a temperature estimate from the average line properties 
formed over the visible hemisphere, and we then adopt this temperature as the 
polar temperature in the calculation of a sequence of rotating star profiles.  
This inconsistency will have only a minor affect on our derived $V\sin i$ 
values because the profile shapes are much more dependent on the Doppler shifts 
caused by rotation than on the adopted temperature of the local atmosphere. 

A similar problem may arise because of our assumption of solar abundances. 
Many He-strong and He-weak stars are known to exist among the B-stars \citep{mat04},
and since our temperature estimation scheme is based on \ion{He}{1} line strengths, 
some of our derived spectral subtype estimates may be incorrect. 
However, even in these cases the assumed profile shapes will probably 
be satisfactory because the Doppler broadening due to rotation is much 
more important for line shape than the assumed chemical abundance.  
We found only one extreme case (NGC~6193 \#17 = CD$-48^\circ11051$) for 
which the \ion{He}{1} line strengths were so strong that none of the solar 
abundance models made a satisfactory fit. 

Finally, we note that our assumption about the spin axis orientation angle has 
a negligible impact on our results.  We calculated the line profiles assuming a fixed inclination 
angle of $i=90^\circ$, but the actual spin axes of the stars point randomly in space. 
In order to determine the consequences of this simplification for the final 
results, we tested our fitting code using rotationally broadened line profiles made 
assuming different inclination angles.  We found that the line profiles change very little 
from $i=90^\circ$ to $i=30^\circ$ for a fixed geometric value of $V\sin i$, 
and for the infrequent cases of smaller inclination angle 
($i < 30^\circ$) the fitting code can still find the correct $V \sin i$ using a 
slightly different spectral type.

%%%%%%%%%%%%%%%%%%%%%%%%%%%%%%%%%%%%%%%%%%%%%%%%%%%%%%%%%%%%%%%

\section{Results and Discussion}                             % Section 4

We obtained $V \sin i$ measurements for a total of 496 stars in 19 young clusters.
Our results for the individual stars in each cluster are listed in Table 3, which
is given in full only in the on-line edition.  The columns list: 
(1) the cluster name; 
(2) WEBDA identification number of the star (coordinates are given 
in the last column for stars with no WEBDA identification);
(3) derived $V \sin i$ averaged over the values obtained from 
\ion{He}{1} $\lambda\lambda 4026, 4387, 4471$, and/or \ion{Mg}{2} $\lambda 4481$; 
(4) the measurement error of $V \sin i$ based on the fit of \ion{He}{1} $\lambda 4471$
and/or \ion{Mg}{2} $\lambda 4481$; 
(5), (6), (7) the radial velocity for each observing night, based on measurements 
of the positions of the \ion{He}{1} lines (note that the targets were observed on three nights
for NGC~2244 and IC~2944, only one night for Trumpler~14 and 16, and on 2 nights for the rest); 
(8) the assigned index for spectral subtype of the star (Table~2); and 
(9) notes for interesting stars, such as double-lined spectroscopic  
binaries (SB2) or stars having unusual line profiles, where the $V \sin i$ measurement 
may not be reliable. For double-lined spectroscopic binary systems, the $V \sin i$ values 
are determined by fitting the dominant component in the double-peaked line profiles.
The digital version of Table~3 also gives the mid-exposure heliocentric Julian dates of 
each observation.  There are several instances where we have added some nearby 
B-type stars to the cluster roster (especially in the case of NGC~2384), and  
although the cluster membership of these stars needs confirmation, we will 
assume in the following that all the stars in Table~3 are cluster members. 

\placetable{tab3}      % Table 3 - V sin i measurements

The radial velocities were obtained by measuring the shift required to 
match the model and observed profiles and then applying a 
correction for Earth's motion in the direction of the star.   
The measurement errors depend 
on the S/N of the individual spectrum but are typically $\pm 10$ km~s$^{-1}$.
However, there are systematic radial velocity errors present 
in several clusters that 
are at least this large and that are manifested in overall differences 
in cluster mean velocity between nights.  Our main interest 
in the velocity data is to identify the radial velocity variable stars 
that are candidate short period spectroscopic binaries.   We adopted a  
limit of a radial velocity shift greater than 30 km~s$^{-1}$ 
between two nights as the criterion for detection of possible binaries. 
However, to avoid any systematic shifts in radial velocity between nights, 
we did not directly use the raw measurements of radial
velocity listed in Table~3 to estimate such variations.  Instead, we calculated the
mean radial velocity for each night and for every cluster first, 
and then we compared the residuals 
(the radial velocity minus the mean) between different nights to find the
binaries using the 30 km~s$^{-1}$ criterion.  This step was done in an iterative
way for each cluster: once we found the binary candidates, we removed them from 
the list and recalculated the mean using the rest of stars and we repeated this
isolation of the binary candidates and calculation of the mean until no more stars 
met the binary star criterion.  The velocity variable stars are identified 
in Table~3 by the notation SB1 in the final column.  The stars that we found 
to be double-lined spectroscopic binaries were also usually velocity variable  
and these are noted as SB2 in Table~3. 

We summarize in Table~4 the results for each of the 19 clusters, which are listed 
in order of increasing age.  The columns here indicate:
(1) the common name for the cluster;
(2) the logarithm of the cluster age from \citet*{lok01}; 
(3) the number $N$ of stars measured; 
(4) the mean projected rotational velocity of the whole sample; 
(5) the number $n$ of stars with a pseudo-spectral subtype of B3 or earlier; 
(6) the mean projected rotational velocity of the subsample of 
those stars of type B3 or earlier; 
(7) the number $n$ of very rapid rotators, i.e., stars with a projected rotational 
velocity greater than 300 km~s$^{-1}$; 
(8) the number $n$ of candidate spectroscopic binary stars; and 
(9) the observatory where the spectra were obtained (see Table~1).
Note that no binary estimates appear for Trumpler 14 and 16 since we 
have only single observations available for these clusters. 
The age determinations from \citet{lok01} are based on fits of 
cluster color--magnitude diagrams made using isochrones from 
\citet{ber94}, and although newer and more accurate estimates 
exist in some cases, we adopt these age determinations here 
because they form a homogeneous set that should provide 
a consistent assignment of ages to the clusters in our sample. 
These age estimates have typical errors of $\pm0.2$~dex 
(see Fig.~5 of \citealt{lok01}). 

\placetable{tab4}      % Table 4 - Cluster Summary Data

We begin our review of these results by considering the properties of 
all the stars in the full sample of clusters.  
The stars of our sample correspond mainly to MK spectral types from O9.5 to B8.5, and
to luminosity classes from III to V.  We plot in Figure~3 the mean $V \sin i$ versus 
pseudo-spectral subtype with the total number of stars in each subtype bin given below. 
These numbers show that our sample 
is biased toward earlier subtypes (luminosity biased).  
The distribution of $V\sin i$ for cluster stars is basically 
flat over this range and appears to similar the classical diagram for 
nearby B-type stars presented by \citet{sle70}.
There is a slight indication that there are 
more fast rotators among the late-B subtypes (B5 - B8), which
is similar to what \citet{bro97} found for stars in Sco~OB2 
and to what \citet*{abt02} found for nearby field stars.
However, the differences between the subtype groups are modest. 
We plot in Figure~4 the cumulative distribution functions 
of $V\sin i$ for three subgroups, 
[O9.5, B1.5], (B1.5, B5.0], and (B5.0, B9.0].
The three curves are all similar to each other  
and their mean $V \sin i$ values are almost same: 
139 km~s$^{-1}$, 154 km~s$^{-1}$, and 151 km~s$^{-1}$ 
for these three groups, respectively.  

\begin{figure}
%\epsscale{0.75}
%\plotone{f3.eps}
\plotfiddle{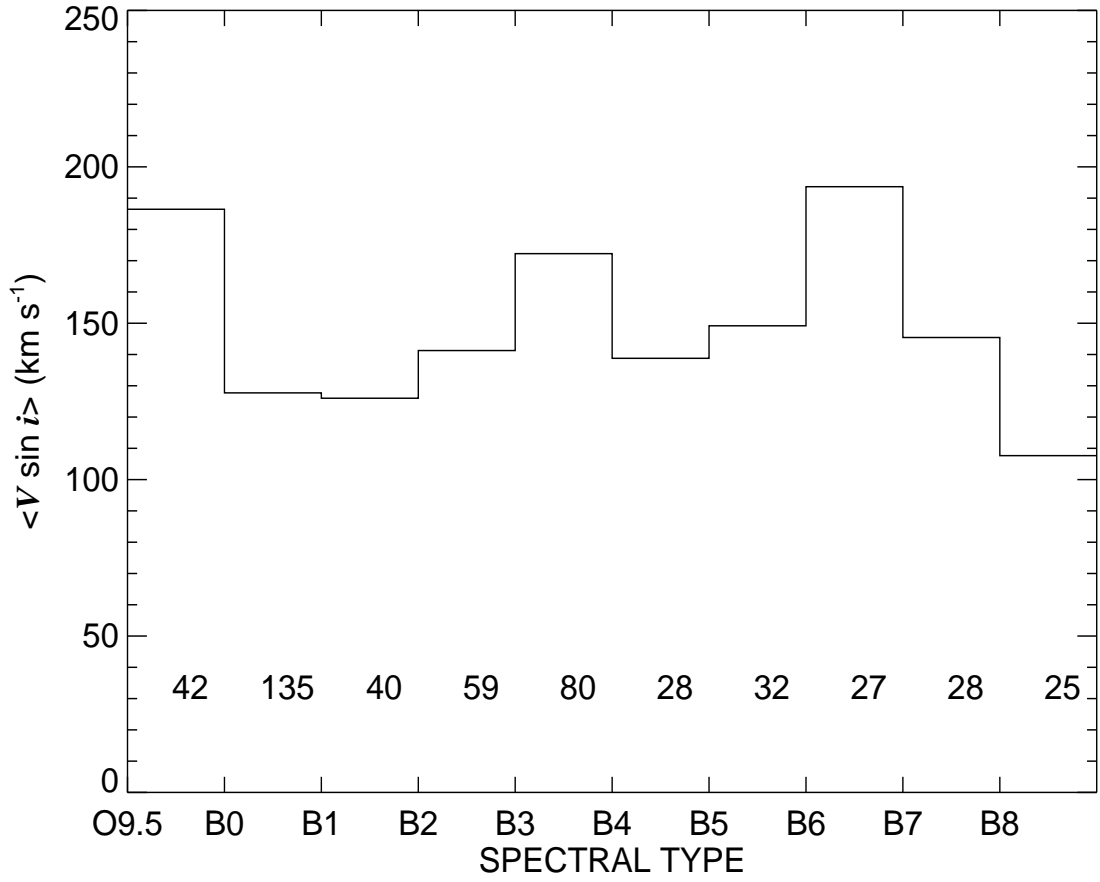}{100pt}{90}{360}{450}{30}{0}
\caption{
Mean $V \sin i$ as a function of pseudo-spectral subtype. 
The number of stars in each subtype bin is given at the bottom.
}
\label{fig3}   
\end{figure}

\placefigure{fig3}     % Figure 3 - Histogram of <V sin i> vs. spectral subtype

\begin{figure}
%\epsscale{0.75}
%\plotone{f4.eps}
\plotfiddle{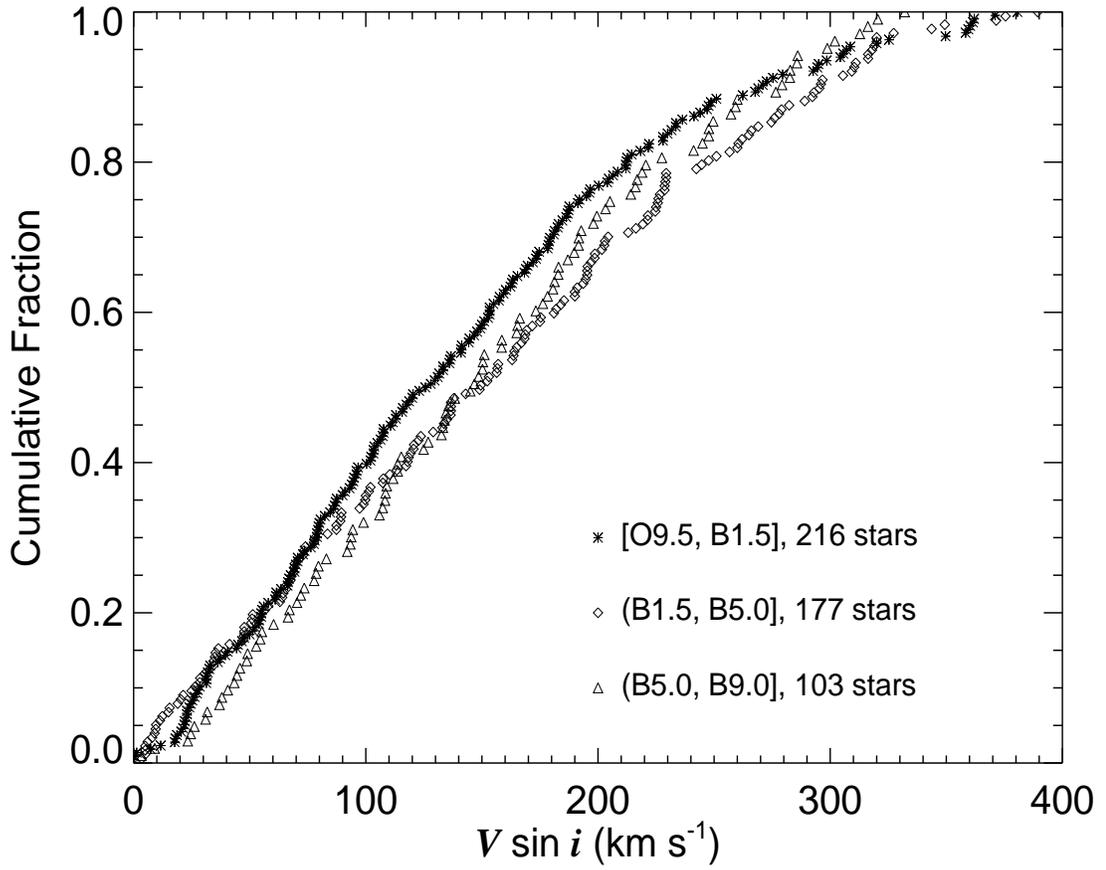}{100pt}{90}{360}{450}{30}{0}
\caption{
The cumulative distribution functions of $V\sin i$ for 
the stars gathered into three groupings of pseudo-spectral subtype. 
The subtype range is given using 
the mathematical notation in which ``['' means
the boundary is included while ``('' means the boundary is not included.}
\label{fig4}
\end{figure}

\placefigure{fig4}     % Figure 4 - CDF of V sin i for 3 spectral subtype groups 

We next compare the projected rotational velocity distributions of cluster and field B-stars. 
A histogram of the distribution of $V \sin i$ for all the measured stars (subtypes O9.5 to B8.5) 
is plotted in Figure~5 ({\it solid line}).  We extracted a subset of similar data
for field stars from \citet{abt02} for stars with subtypes of B8.5 or earlier 
and luminosity classes between III and V (737 stars), and we plot the field star histogram 
also in Figure~5 ({\it dashed line}).  The comparison shows that there are more 
slow rotators among the field B stars than is the case for the cluster stars. 
The mean projected rotational velocities are $113\pm3$ km~s$^{-1}$ 
and $148\pm4$ km~s$^{-1}$ for the field and cluster stars, respectively 
(where the quoted errors are the standard deviation of the mean).
Part of this difference may be due to the better spectral resolution of 
the spectra used by \citet{abt02} (better able to discern the sharpest-lined
stars) and due to the presence of unrecognized binaries in our sample 
(with blended and broadened line profiles).  It is also possible 
that some real physical causes are responsible for the difference. 
For example, \citet*{str05} argue that stars formed in denser 
environments (clusters) appear to rotate faster than field stars. 
It is also possible that the field stars may
represent a somewhat older population (stars from dispersed clusters)
so that the difference reflects evolutionary spin-down (see below).

\placefigure{fig5}     % Figure 5 - V sin i histogram for cluster and field stars

The cumulative distribution function of $V \sin i$ for whole sample is plotted
in Figure~6 ({\it top panel}).  We made a 4th order polynomial fit to the 
cumulative distribution function (shown as a {\it dotted line} 
in the {\it top panel}), and we then obtained an estimate of the 
$V \sin i$ probability distribution from the derivative of the fit curve 
(shown as the {\it dotted line} in {\it bottom panel}).
We next calculated the underlying probability distribution for
the stellar equatorial velocity $V_{\rm eq}$ ({\it dashed line} in the {\it bottom panel})
from the $V \sin i$ distribution using the deconvolution algorithm 
from \citet{luc74}.  The $V \sin i$ distribution of our sample peaks 
near 100 km~s$^{-1}$ while the $V_{\rm eq}$ distribution peaks near 200 km~s$^{-1}$.

\placefigure{fig6} % Figure 6 Lucy deconvolution

We expect that the rotational velocity distributions of single and 
close binary stars may differ, since binary systems may experience 
either spin-down through tidal braking or a spin-up from mass transfer. 
There is evidence of rotational and orbital synchronization 
among binaries in the field population (\citealt*{wol82}; \citealt{abt02}).
There are 94 variable radial velocity stars ($21\%$) among the 
454 stars in our sample with more than one radial velocity measurement
that meet the spectroscopic binary criterion we adopted 
($\Delta V_r >~30$ km~s$^{-1}$).  The mean $V \sin i$ 
of the candidate binary group is $165\pm9$ km~s$^{-1}$ compared
to a mean of $144\pm5$ km~s$^{-1}$ for the remaining radial velocity 
constant stars.   This is a somewhat surprising result since 
synchronously rotating binary stars among the field population have
a lower mean $V\sin i$ than single stars \citep{abt02}. 
We wondered if the larger radial velocity measurement errors associated 
with the rapid rotators (which have broad and shallow lines) led 
to the accidental inclusion of some rapid rotators in the candidate 
binary group.  Consequently, we selected a second group of binary 
candidates using a more stringent radial velocity variation 
criterion ($\Delta V_r >~50$ km~s$^{-1}$).  The adoption of 
this criterion reduces the number of candidate binaries to 49, and 
the mean $V \sin i$ for this group is even higher, 
$173\pm15$ km~s$^{-1}$.  The cumulative distribution functions for 
all three groups of stars (single stars, binary stars from the  
30 km~s$^{-1}$ criterion, and binary stars from the 50 km~s$^{-1}$ criterion) 
are plotted in Figure~7.  The Kolmogorov-Smirnov test (KS, hereafter)
indicates that there is only a $1\%$ probability that the the single and 
binary star ($\Delta V_r >~30$ km~s$^{-1}$) data are drawn from the same distribution
(or a $3\%$ probability for the binary group selected using the tighter constraint, 
$\Delta V_r >~50$ km~s$^{-1}$).  Thus, the difference appears to
be significant. One possible explanation could be that
our binary group includes many cases
in which a faint or only partially resolved spectral line from the companion
causes the composite profile to appear broader and leads to a higher
value of $V\sin i$ for such binaries.  On the other hand, 
some of these systems may have experienced a spin-up of the 
mass gainer due to mass transfer. 
Follow up spectroscopic investigations of the candidate binaries
will be required to determine the precise nature
of their projected rotational velocities.

\placefigure{fig7}     % Figure 7 - CDFs for single and binary stars

We next turn to the question of temporal changes in the rotational 
velocity distributions by considering the cluster ages.
One way to show the possible evolutionary effects of rotation 
for B-stars in these clusters is by plotting the $V \sin i$ 
cumulative distribution functions for clusters of different ages.
Since the effects of rotation are expected to occur on time scales 
comparable to the main sequence lifetime \citep{heg00,mey00} and 
since our sample stars (O9 - B9) have dramatically different MS durations,
we decided to divide our sample into two groups,
high mass ($\geq 9 M_\odot$) and low mass ($< 9 M_\odot$).
The main sequence lifetime of a $9 M_\odot$ star is $\log {\rm age} = 7.4$
\citep{sch92}, and since most of the clusters we observed are younger 
than this, only stars in the high mass group in our sample are expected to display
evolutionary changes.  For each cluster, we first derived the absolute
magnitude $M_V$ of a $9 M_\odot$ star at the cluster age using theoretical photometry
data from \citet{lej01}.  We next transformed this into an apparent magnitude $m_V$ using
the reddening-corrected distance modulus of the cluster.  Then we assumed 
that all cluster stars with a magnitude $\leq m_V$ belong to the high mass group,
while the fainter stars belong to the low mass group.  Ideally, we would like 
to have a sufficient number of stars of each group in every cluster 
for statistical analysis, but since we observed fewer than 20 stars 
in about half of the clusters, this
kind of division is impractical.  Thus, we grouped
clusters with similar ages together into three age bins which 
have adequate numbers of stars for the cumulative function plots.
The cumulative distribution functions of $V\sin i$ for the
different age bins in both high and low mass groups are shown in Figure~8,
arranged in panels of 2 columns for the mass groups
by 3 rows for the various age bins.  Every panel shows two distributions,
the cumulative function for all the stars in 
the corresponding mass group ({\it filled dots}),
which appears the same in each panel of a given column, and
the cumulative function for the specific age bin in this mass group
({\it diamonds}).   We also list in each panel the KS probability
that the all-age and specific-age binned $V\sin i$ data are drawn from the
same distribution.

The expected spin-down is not immediately evident in Figure~8, and, in fact, both 
mass groups show evidence of a possible spin-up between the first two age bins 
(perhaps followed by a spin-down in the last age bin).  
The mean $V \sin i$ values for the three age bins of the high mass group, from young
to old, are $119\pm11$, $136\pm8$, and $126\pm19$ km~s$^{-1}$ 
(based upon 67, 111, and 22 stars, respectively).  The KS probabilities 
that the functions are drawn from the same population are 
0.28 between the functions shown in the top and middle panels and 
0.85 between the functions shown in the middle and bottom panels, respectively.  
The mean $V \sin i$ values of the three age bins for the low mass group, from young
to old, are $156\pm7$, $174\pm12$, and $152\pm12$ km~s$^{-1}$ 
(from samples of 146, 73, and 62 stars, respectively).  The KS probabilities
are 0.20 between the functions in the top and middle panels and 
0.11 between the functions in the middle and bottom panels, respectively.  
Both groups show that more rapid rotators exist in
the middle age bin, which contains clusters somewhat older than 10~Myr.  

\placefigure{fig8}     % Figure 8 - CDFs for different age bins

We plot the mean $V \sin i$ of each cluster against the cluster age in Figures~9 
(high mass group) and 10 (low mass group), where we omit those clusters with 
fewer than 6 measurements.   We also plot in Figure~9 (high mass group)  
theoretical evolutionary tracks of the mean $V \sin i$ for single stars 
with masses of $9$ and $12 M_\odot$ based on the results from \citet{mey00}.
We assumed 
that all clusters begin with a distribution of stellar rotational velocities given by the
$V_{\rm eq}$ distribution in Figure~6. Then we used the \citet{mey00} model results to calculate 
how each $V_{\rm eq}$ point is transformed to a lower velocity at each time step 
in order to determine a new mean $V_{\rm eq}$ for that time step.  The physical mean
$<V_{\rm eq}>$ was then transformed into the projected mean $<V \sin i>$ by multiplying a factor
of $\pi/4$ (assuming that the spin axes of stars point randomly in space).  
With a few exceptions, most of the cluster means in Figure~9 appear to fall within  
the region defined by the theoretically predicted spin down tracks.  However, 
both Figures~9 and 10 confirm the impression from Figure~8 that clusters
in the age range $7.0 < \log {\rm age} < 7.15$ seem to contain more rapid rotators. 

\placefigure{fig9}     % Figure 9 - mean V sin i vs. age for high mass stars 

\placefigure{fig10}    % Figure 10 - mean V sin i vs. age for low mass stars

We expect that rapid rotators originate in three ways: 
1) stars that are born as very rapid rotators and maintain 
their spin throughout the MS phase; 
2) mass transfer in a binary system can turn the mass gainer into a very rapid rotator; and
3) the core contraction of a star at the TAMS can cause a spin-up. 
The last possibility can be investigated by considering the 
the locations of the rapid rotators in the cluster HRD.  
%If they spin up at the TAMS as predicted by \citet{heg00,mey00}, then we might expect 
%to find these fast rotators located near the TAMS region in the HRD. 
%Models of the evolution of massive rotating stars \citep{heg00,mey00} 
%predict that stars experience a short duration spin up episode 
%caused by interior core contraction as the star approaches the TAMS.
%If correct, then we might expect to find these fast rotators 
%located near the TAMS region in the HRD.  
We selected four clusters that each contain a significant number of 
rapidly rotating stars with $V \sin i > 260$ km~s$^{-1}$. 
The two clusters with the fastest mean $V\sin i$ 
are NGC~3293 and Berkeley~86, and both of these have an age near 10~Myr. 
We found that 9 of the 23 stars we observed in NGC~3293 have 
$V \sin i > 260$ km~s$^{-1}$ (and one additional fast rotator 
was identified by \citealt{bal75b}).  
There are 4 of 17 stars with $V \sin i > 260$ km~s$^{-1}$ 
among those observed in the cluster Berkeley~86.  
We also considered two older clusters, NGC~4755 and NGC~457, 
for which we identified 5 of 33 stars and 
3 of 19 stars, respectively, with $V \sin i > 260$ km~s$^{-1}$.
We plot the positions of all these fast rotators 
in color-magnitude diagrams in Figure~11.  Only two of these
stars are found near the TAMS in the region where single star models 
predict that the spin-up may occur (NGC~3293 \#19 and Berkeley~86 \#9). 
We caution that even the positions 
of these two stars may be artificially shifted to redder colors by 
rotation and gravity darkening and so they may not actually be TAMS
objects.   The rapid rotators appear to be located throughout the 
band from the ZAMS to the TAMS in the HRD much as are the rapidly rotating 
Be stars \citep{zor04,mcs05}.  This suggests that the fast rotators observed 
in these clusters are not generally the result of the predicted core adjustment 
that occurs at TAMS.  

\placefigure{fig11}     % Figure 11 - Cluster HRDs with positions of rapid rotators 

However, clusters with an age of $\approx 10$~Myr 
probably contain close binary systems that are old enough for the 
evolutionary expansion and Roche lobe overflow to occur, and 
the mass gainers could be spun up to become rapid rotators 
(see the study of the Algol binary RY~Per by \citealt{bar04}). 
We find evidence for only two candidate binaries among these 
rapid rotators (NGC~3293 \#83 = CPD$-57^\circ$3501, and
NGC~457 \#43), but post-mass transfer
binaries are probably not easily detected in our survey since they 
probably have long orbital periods and small radial velocity semi-amplitudes
(see, for example, the study of the Be binary $\phi$~Per by \citealt{gie98}). 
Models of the binary spin-up process \citep{pol91,van97} suggest 
that most of the mass gainer stars will appear among the more 
massive stars near the top of the main sequence.   It will 
be important to compare our results on the rotational velocities 
of the brighter cluster members with investigations of the 
rotational properties of the fainter, lower mass population 
in order to test this prediction. 

Our approach in this paper was to group individual stars 
according to cluster membership and age to study 
how the rotational velocity distribution varies with time. 
However, since the predicted evolutionary time scales vary 
with stellar mass, we are probably losing some information 
on the rotational evolution by averaging over samples with 
a significant range in stellar mass.  In the next paper in 
this series \citep{hua06}, we will present a scheme we have developed to 
determine the effective temperature, gravity, and helium 
abundance for each of our targets, and 
we will present additional evidence for the evolutionary 
spin down that is predicted to occur for massive rotating stars. 

%%%%%%%%%%%%%%%%%%%%%%%%%%%%%%%%%%%%%%%%%%%%%%%%%%%%%%%%%%%%%%%

\acknowledgments

We are grateful to the KPNO and CTIO staffs and especially 
Diane Harmer and Roger Smith for their help in making these
observations possible.  We thank Richard Townsend
and Paul Wiita for their very helpful comments.
We also thank Ivan Hubeny and 
Thierry Lanz for their assistance with the TLUSTY and SYNSPEC codes.   
This material is based on work supported by the National Science 
Foundation under Grant No.~AST-0205297. 
Institutional support has been provided from the GSU College
of Arts and Sciences and from the Research Program Enhancement
fund of the Board of Regents of the University System of Georgia,
administered through the GSU Office of the Vice President for Research.
We gratefully acknowledge all this support.

%%%%%%%%%%%%%%%%%%%%%%%%%%%%%%%%%%%%%%%%%%%%%%%%%%%%%%%%%%%%%%%

% References

%%%%%%%%%%%%%%%%%%%%%%%%%%%%%%%%%%%%%%%%%%%%%%%%%%%%%%%%%%%%%%%

% Tables

\clearpage

% Table 1 - Observational Parameters

\begin{deluxetable}{lcc}
\tablewidth{0pc}
\tablecaption{Hydra Observing Runs\label{tab1}}
\tablehead{
\colhead{Parameter} &  
\colhead{WIYN 3.5 m} &
\colhead{CTIO 4.0 m} }
\startdata
UT Dates                   \dotfill & 2000 Nov. 13 -- 16 & 2001 Feb. 9 -- 12   \\
Fiber diameter ($\mu$m)    \dotfill &  310      &  300   \\
Camera name                \dotfill & Simmons   & Bench Schmidt \\
Camera focal length (mm)   \dotfill & 381       &  400   \\
Grating name               \dotfill & 1200@28.7 & KPGLD  \\
Grating grooves mm$^{-1}$  \dotfill & 1200      &   790  \\
Grating blaze angle (deg)  \dotfill &  28.7     &  19.6  \\
Order                      \dotfill &  2        &   2    \\
Order sorting filter       \dotfill &  CuSO4    &  BG-39 \\
CCD detector               \dotfill & SITe $2\times2$K & SITe $2\times4$K \\
CCD pixel size ($\mu$m)    \dotfill &   24      &   15   \\
Reciprocal dispersion (\AA ~pixel$^{-1}$) \dotfill & 0.252  &  0.440 \\
Wavelength range (\AA )    \dotfill & 3983 -- 4501 & 3878 -- 4781 \\ 
Comparison source          \dotfill &  CuAr     &  He-Ne-Ar  \\
FWHM at 4026 \AA ~(\AA )   \dotfill &  0.74     &  1.74  \\
FWHM at 4387 \AA ~(\AA )   \dotfill &  0.83     &  1.44  \\
FWHM at 4471 \AA ~(\AA )   \dotfill &  0.85     &  1.42  \\
\enddata
\end{deluxetable}

%%%%%%%%%%%%%%%%%%%%%%%%%%%%%%%%%%%%%%%%%%%%%%%%%%%%%%%%%%%%%%%
\clearpage

% Table 2 - Model Stellar Parameters 

\begin{deluxetable}{lcccc}
\tablewidth{0pc}
\tablecaption{ZAMS Model Star Parameters
%\tablenotemark{*} 
\label{tab2}}
\tablehead{
\colhead{ } &
\colhead{Pseudo-} &
\colhead{ } &
\colhead{ } &
\colhead{ } \\
\colhead{ } &
\colhead{Spectral} &
\colhead{$T_{\rm eff}$} &
\colhead{$M$} &
\colhead{$R$} \\
\colhead{Index} &
\colhead{Subtype} &
\colhead{(kK)} &
\colhead{($M_\odot$)} &
\colhead{($R_\odot$)}}
\startdata
\phn 0\dotfill & O9.5 V &    34.9 &     20.0 & 5.8 \\
\phn 1\dotfill & O9.8 V &    32.6 &     17.0 & 5.3 \\
\phn 2\dotfill & B0.0 V &    31.1 &     15.0 & 5.0 \\
\phn 3\dotfill & B0.3 V &    29.1 &     12.6 & 4.6 \\
\phn 4\dotfill & B0.5 V &    27.8 &     11.0 & 4.3 \\
\phn 5\dotfill & B1.0 V &    25.5 & \phn 9.5 & 4.0 \\
\phn 6\dotfill & B1.5 V &    23.7 & \phn 8.3 & 3.7 \\
\phn 7\dotfill & B2.0 V &    22.0 & \phn 7.0 & 3.3 \\
\phn 8\dotfill & B2.5 V &    20.3 & \phn 6.3 & 3.1 \\
\phn 9\dotfill & B3.0 V &    18.7 & \phn 5.5 & 2.9 \\
    10\dotfill & B3.5 V &    17.8 & \phn 5.1 & 2.9 \\
    11\dotfill & B4.0 V &    17.0 & \phn 4.8 & 2.8 \\
    12\dotfill & B4.5 V &    16.2 & \phn 4.4 & 2.8 \\
    13\dotfill & B5.0 V &    15.4 & \phn 4.0 & 2.7 \\
    14\dotfill & B5.5 V &    14.9 & \phn 3.8 & 2.6 \\
    15\dotfill & B6.0 V &    14.4 & \phn 3.6 & 2.5 \\
    16\dotfill & B6.5 V &    13.9 & \phn 3.4 & 2.5 \\
    17\dotfill & B7.0 V &    13.5 & \phn 3.3 & 2.4 \\
    18\dotfill & B7.5 V &    13.0 & \phn 3.2 & 2.3 \\
    19\dotfill & B8.0 V &    12.6 & \phn 2.9 & 2.2 \\
    20\dotfill & B8.5 V &    11.8 & \phn 1.8 & 2.1 \\
\enddata
%\tablenotetext{*}{O9.5-B5.0: based on Table 3 in Hanson 1997\\ 
%B5.0 - B8.5: based on Appendix B in Gray, 1992}
\end{deluxetable}

%%%%%%%%%%%%%%%%%%%%%%%%%%%%%%%%%%%%%%%%%%%%%%%%%%%%%%%%%%%%%%%
\clearpage

% Table 3 - V sin i results

\begin{deluxetable}{lcccccccc}
\tabletypesize{\footnotesize}
\tablewidth{0pc}
\tablecaption{Projected Rotational Velocities of B Stars in 
19 Open Clusters\label{tab3}\tablenotemark{a}}
\tablehead{
\colhead{  } &  
\colhead{  } &  
\colhead{  } &
\colhead{  } &
\colhead{  } &
\colhead{  } &
\colhead{  } &  
\colhead{Pseudo-} &
\colhead{  }\\

\colhead{  } &  
\colhead{  } &  
\colhead{  } &
\colhead{  } &
\colhead{  } &
\colhead{  } &
\colhead{  } &  
\colhead{Spectral} &
\colhead{  }\\

\colhead{Cluster} &  
\colhead{WEBDA} &
\colhead{$V \sin i$} &
\colhead{$\delta V \sin i$} &
\colhead{$V_r (N1)$} &
\colhead{$V_r (N2)$} &
\colhead{$V_r (N3)$} &
\colhead{Subtype} &
\colhead{   }\\

\colhead{Name} &  
\colhead{Number} &
\colhead{(km s$^{-1}$)} &
\colhead{(km s$^{-1}$)} &
\colhead{(km s$^{-1}$)} &
\colhead{(km s$^{-1}$)} &
\colhead{(km s$^{-1}$)} &
\colhead{Index} &
\colhead{Notes}}

\startdata
Ber 86\dotfill  & 1 &  184 & \phn 7 & \phs  30.3  & \phs    30.8  & \nodata & 1.3 & \nodata \\
Ber 86\dotfill  & 3 &  192 &     10 &     $-24.6$ & \phs\phn 7.5  & \nodata & 4.0 & SB2     \\
Ber 86\dotfill  & 4 &  178 & \phn 8 & \phs  17.2  & \phs    45.2  & \nodata & 1.0 & SB2     \\
Ber 86\dotfill  & 9 &  362 &     21 & \phn $-7.6$ &       $-15.7$ & \nodata & 3.3 & \nodata \\
\enddata
\tablenotetext{a}{The full version of this table appears in the on-line edition.}
\end{deluxetable}

%%%%%%%%%%%%%%%%%%%%%%%%%%%%%%%%%%%%%%%%%%%%%%%%%%%%%%%%%%%%%%%
\clearpage

% Table 4 - Cluster Summary
\begin{deluxetable}{lcccccccc}
\tabletypesize{\footnotesize}
%\rotate
\tablewidth{0pc}
\tablecaption{Summary Data for 19 Open Clusters \label{tab4}}
\tablehead{
\colhead{  } &
\colhead{  } &
\colhead{  } &
\colhead{$<V \sin i>$  } &
\colhead{  } &
\colhead{$<V \sin i>$  } &
\colhead{$n$  } &
\colhead{  } &
\colhead{  } \\

\colhead{    } &
\colhead{log~age} &
\colhead{   } &
\colhead{(All)  } &
\colhead{$n$} &
\colhead{($\leq$ B3)   } &
\colhead{($V \sin i >$ } &
\colhead{$n$ } &
\colhead{  } \\

\colhead{Name    } &
\colhead{(yr) } &
\colhead{$N$} &
\colhead{(km s$^{-1}$) } &
\colhead{($\leq$ B3)} &
\colhead{(km s$^{-1}$) } &
\colhead{ 300 km s$^{-1}$)\tablenotemark{a}    } &
\colhead{(Binary)\tablenotemark{a}    } &
\colhead{Obs.}}

\startdata
NGC 6193   \dotfill &    6.78 &     20 &     155 & 15     &     134 & 2 (10)    & \phn 4 (20)    & CTIO \\
Trumpler 16\dotfill &    6.79 &     35 &     135 & 27     &     150 & 2 (6)\phn & \nodata        & CTIO \\
IC 1805    \dotfill &    6.82 &     30 &     145 & 25     &     147 & 1 (3)\phn & \phn 9 (30)    & WIYN \\
IC 2944    \dotfill &    6.82 &     38 &     136 & 25     &     127 & 2 (5)\phn & \phn 5 (13)    & CTIO \\
Trumpler 14\dotfill &    6.83 & \phn 6 &     109 & \phn 2 &     178 & 0 (0)\phn & \nodata        & CTIO \\
NGC 2244   \dotfill &    6.90 &     41 &     168 & 15     &     153 & 5 (12)    &     16 (39)    & WIYN \\
NGC 2384   \dotfill &    6.90 &     15 & \phn 91 & \phn 8 & \phn 83 & 0 (0)\phn & \phn 1 (7)\phn & CTIO \\
NGC 2362   \dotfill &    6.91 &     28 &     161 & 11     &     141 & 1 (4)\phn & \phn 4 (14)    & CTIO \\
NGC 3293   \dotfill &    7.01 &     23 &     184 & 20     &     178 & 5 (22)    & \phn 2 (9)\phn & CTIO \\
NGC 884    \dotfill &    7.03 &     57 &     149 & 48     &     140 & 4 (7)\phn &     14 (25)    & WIYN \\
NGC 1502   \dotfill &    7.05 &     18 &     166 & 12     &     183 & 2 (11)    & \phn 5 (28)    & WIYN \\
NGC 869    \dotfill &    7.07 &     55 &     121 & 49     &     113 & 2 (4)\phn &     14 (25)    & WIYN \\
NGC 2467   \dotfill &    7.10 &     14 &     150 & \phn 4 &     167 & 1 (7)\phn & \phn 2 (14)    & CTIO \\
Berkeley 86\dotfill &    7.12 &     17 &     194 & 15     &     181 & 3 (18)    & \phn 1 (6)\phn & WIYN \\
IC 2395    \dotfill &    7.22 &     16 &     113 & \phn 3 & \phn 94 & 0 (0)\phn & \phn 2 (13)    & CTIO \\
NGC 4755   \dotfill &    7.22 &     33 &     141 & 18     &     151 & 1 (3)\phn & \phn 3 (9)\phn & CTIO \\
NGC 7160   \dotfill &    7.28 &     16 &     183 & \phn 5 &     177 & 2 (13)    & \phn 4 (25)    & WIYN \\
NGC 457    \dotfill &    7.32 &     19 &     150 & 10     &     173 & 1 (5)\phn & \phn 6 (32)    & WIYN \\
NGC 2422   \dotfill &    7.86 &     15 &     143 & \phn 1 &     245 & 1 (7)\phn & \phn 2 (13)    & WIYN \\
\enddata
\tablenotetext{a}{The number in parentheses is the percentage.}
\end{deluxetable}
%%%%%%%%%%%%%%%%%%%%%%%%%%%%%%%%%%%%%%%%%%%%%%%%%%%%%%%%%%%%%%%%
% Figures

\clearpage

% Figure 1

%%%%%%%%%%%%%%%%%%%%%%%%%%%%%%%%%%%%%%%%%%%%%%%%%%%%%%%%%%%%%%%

\clearpage

% Figure 2

%%%%%%%%%%%%%%%%%%%%%%%%%%%%%%%%%%%%%%%%%%%%%%%%%%%%%%%%%%%%%%%

\clearpage

% Figure 3

%%%%%%%%%%%%%%%%%%%%%%%%%%%%%%%%%%%%%%%%%%%%%%%%%%%%%%%%%%%%%%%

\clearpage

% Figure 4

%%%%%%%%%%%%%%%%%%%%%%%%%%%%%%%%%%%%%%%%%%%%%%%%%%%%%%%%%%%%%%%

\clearpage

% Figure 5
\begin{figure}
%\epsscale{0.75}
%\plotone{f5.eps}
\plotfiddle{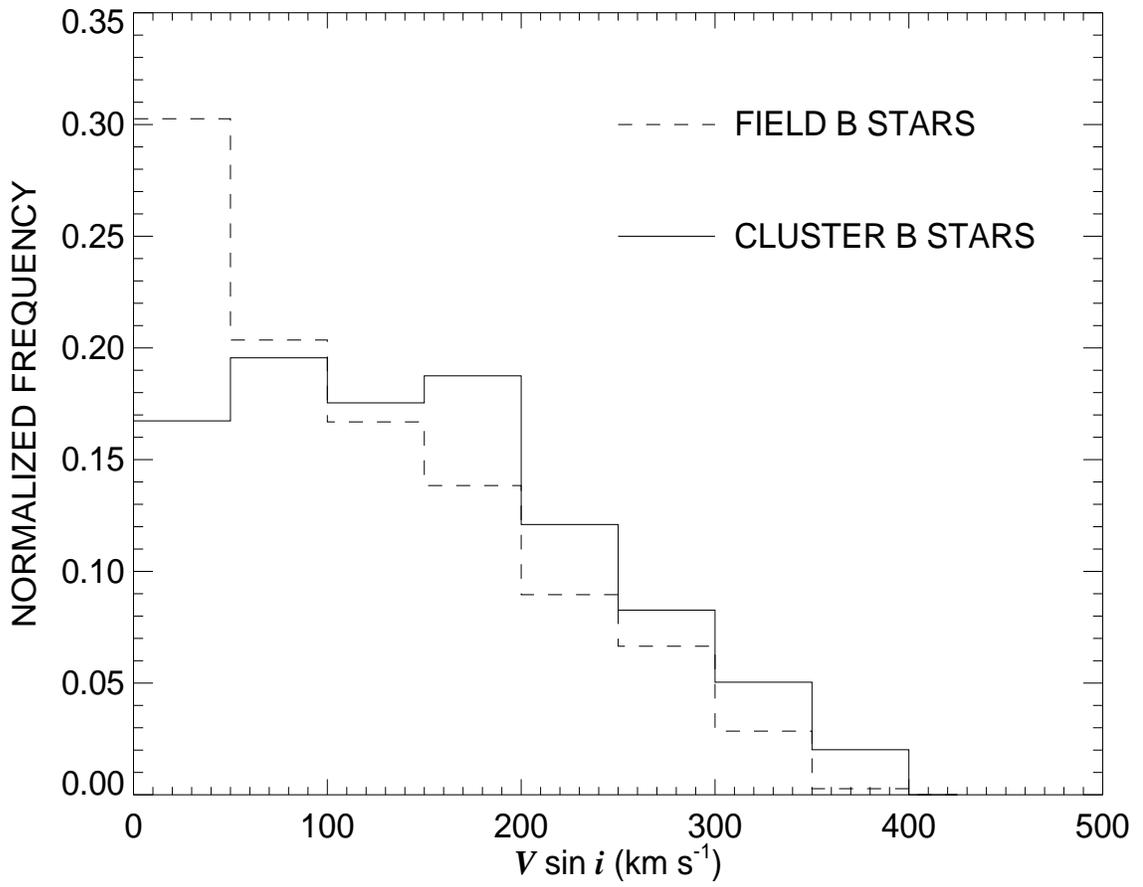}{100pt}{90}{360}{450}{30}{0}
\caption{
Histograms of $V \sin i$ for field stars \citep{abt02}
({\it dashed line}) and for cluster stars ({\it solid line}).
}
\label{fig5}   
\end{figure}

%%%%%%%%%%%%%%%%%%%%%%%%%%%%%%%%%%%%%%%%%%%%%%%%%%%%%%%%%%%%%%%

\clearpage

% Figure 6
\begin{figure}
\epsscale{0.75}
\plotone{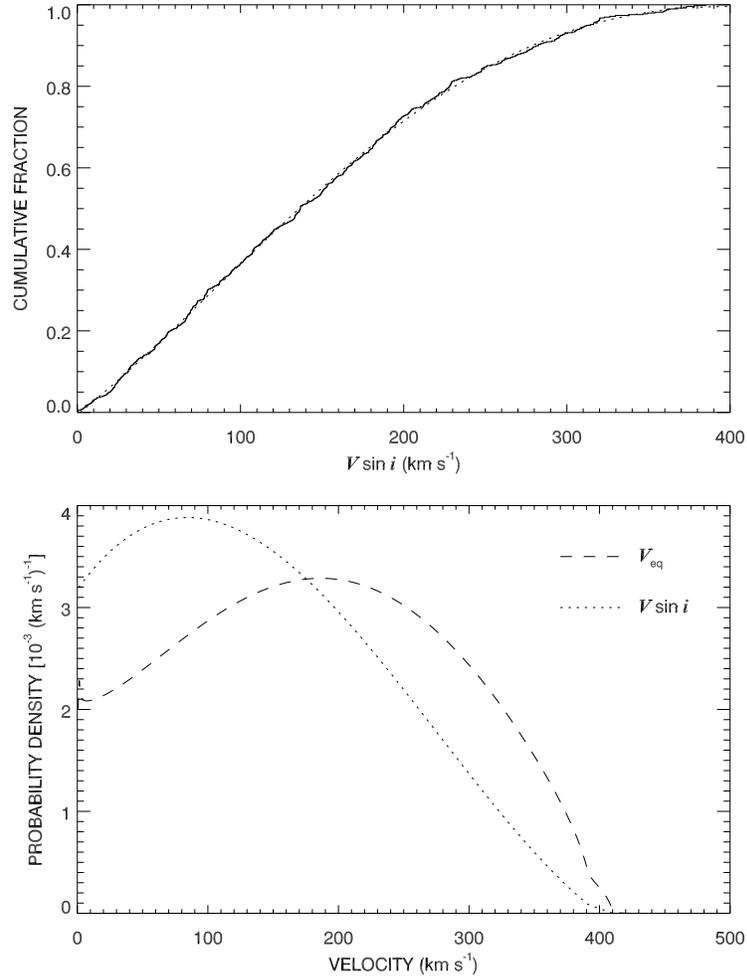}
\caption{
{\it Top panel:} the cumulative distribution function of 
$V \sin i$ for the entire sample of 496 stars ({\it solid line}) 
and its polynomial fit ({\it dotted line}). {\it Bottom panel:} the
$V \sin i$ distribution derived from the polynomial fit ({\it dotted line})
and the associated distribution of equatorial velocity $V_{\rm eq}$ ({\it dashed line}).
}
\label{fig6}
\end{figure}

%%%%%%%%%%%%%%%%%%%%%%%%%%%%%%%%%%%%%%%%%%%%%%%%%%%%%%%%%%%%%%%

\clearpage

% Figure 7
\begin{figure}
%\epsscale{0.75}
%\plotone{f7.eps}
\plotfiddle{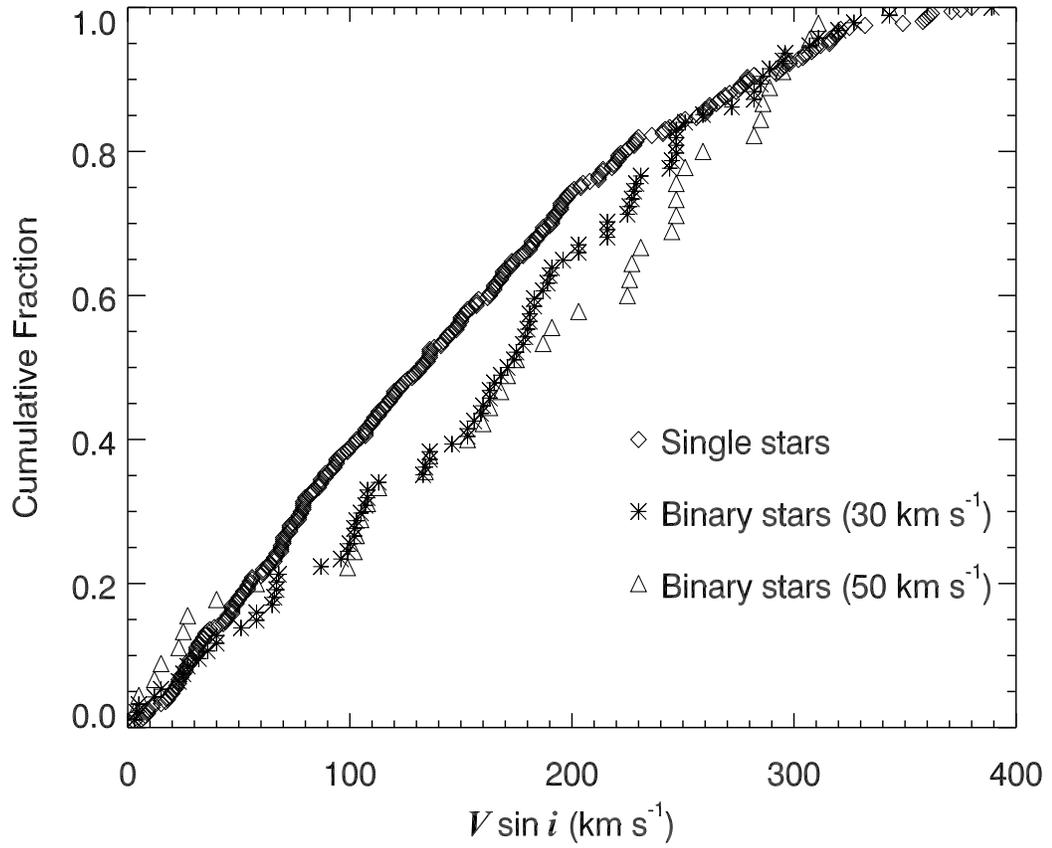}{100pt}{90}{360}{450}{30}{0}
\caption{
The cumulative distribution functions of $V\sin i$ for single 
and candidate binary stars (according to two criteria for 
the minimum radial velocity range).
}
\label{fig7}
\end{figure}  

%%%%%%%%%%%%%%%%%%%%%%%%%%%%%%%%%%%%%%%%%%%%%%%%%%%%%%%%%%%%%%%

\clearpage

% Figure 8
\begin{figure}
\epsscale{0.8}
\plotone{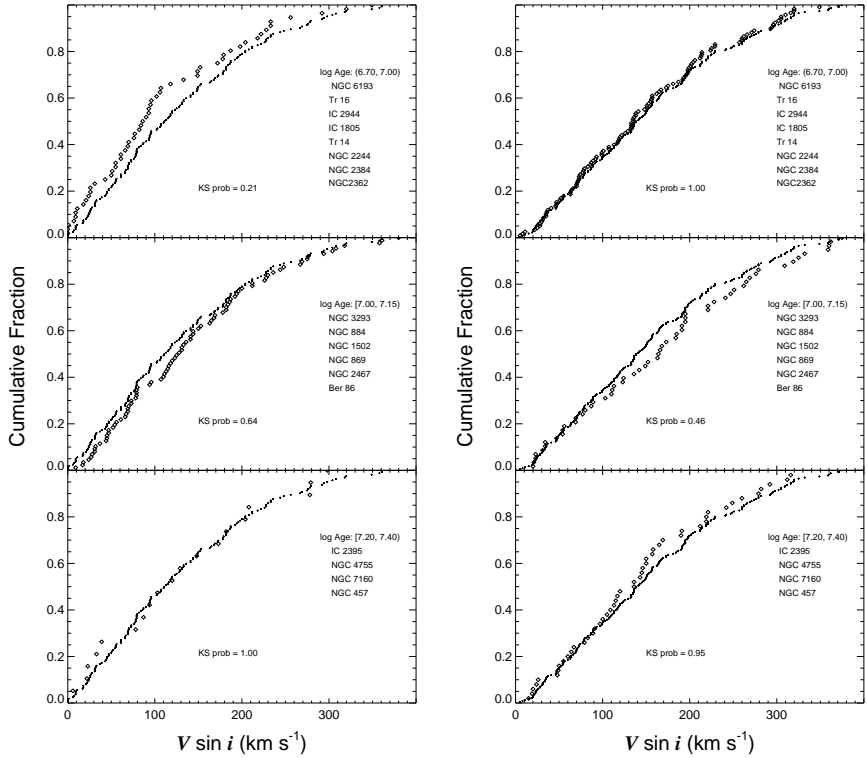}
\caption{
The cumulative distribution functions of $V\sin i$ 
for the subgroups of high mass ($M \geq 9 M_\odot$, {\it left} column) 
and low mass ($M < 9 M_\odot$, {\it right} column) stars at different ages
(increasing from top to bottom).
Each panel has two curves: one is the function for the 
specific-age subgroup ({\it diamonds}) while the other is the function  
for all the stars of the three panels together ({\it filled circles}).
The KS probability that the two samples are drawn from the
same parent distribution is listed in each case. 
}
\label{fig8}   
\end{figure}

%%%%%%%%%%%%%%%%%%%%%%%%%%%%%%%%%%%%%%%%%%%%%%%%%%%%%%%%%%%%%%%

\clearpage  

% Figure 9
\begin{figure}
%\epsscale{0.8}
%\plotone{f9.eps}
\plotfiddle{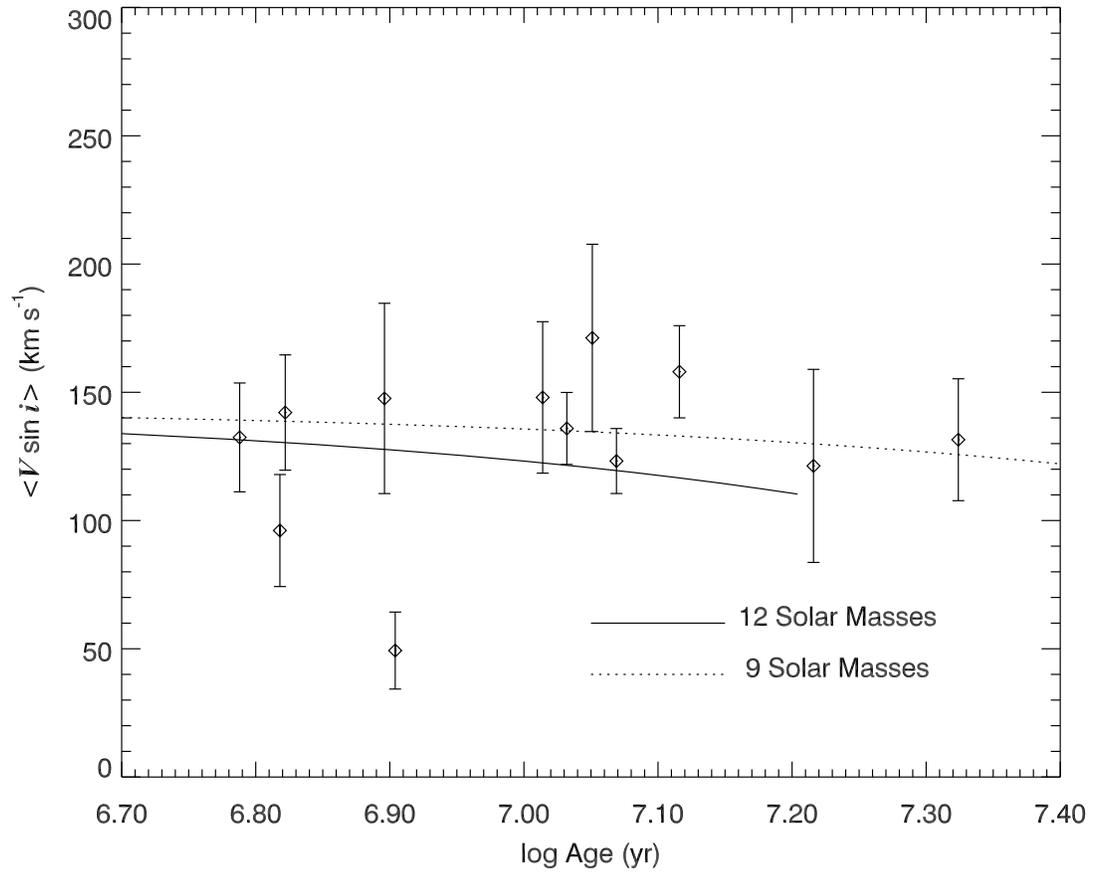}{100pt}{90}{360}{450}{30}{0}
\caption{
Mean $V \sin i$ for high mass stars ($M \geq 9 M_\odot$)
in young clusters plotted against logarithm of cluster age. The solid
line and dotted line show the prediction from \citet{mey00}
for an ensemble of $12 M_\odot$ and $9 M_\odot$ stars, respectively.
}
\label{fig9}
\end{figure}

%%%%%%%%%%%%%%%%%%%%%%%%%%%%%%%%%%%%%%%%%%%%%%%%%%%%%%%%%%%%%%%

\clearpage

% Figure 10
\begin{figure}
%\epsscale{0.8}
%\plotone{f10.eps}
\plotfiddle{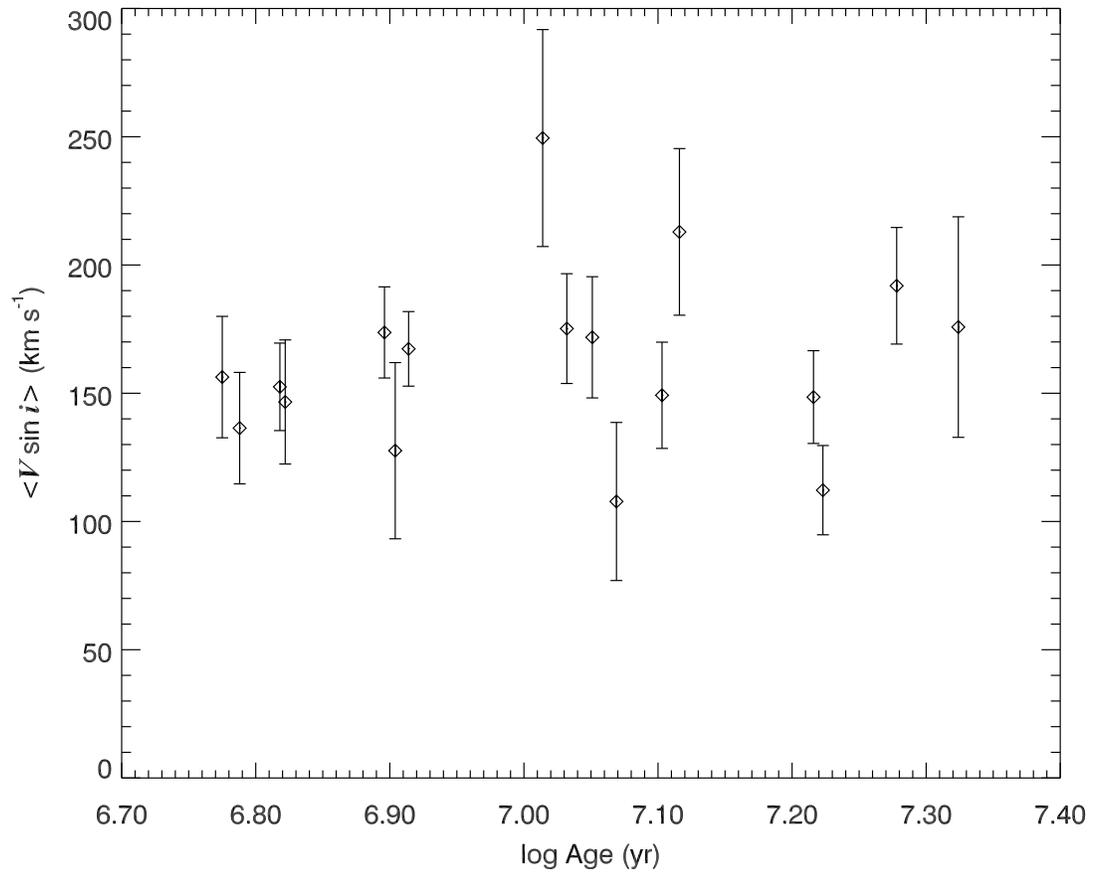}{100pt}{90}{360}{450}{30}{0}
\caption{
Mean $V \sin i$ for stars ($M < 9 M_\odot$)
in young clusters plotted against logarithm of cluster age. 
}
\label{fig10}
\end{figure}

%%%%%%%%%%%%%%%%%%%%%%%%%%%%%%%%%%%%%%%%%%%%%%%%%%%%%%%%%%%%%%%

\clearpage

% Figure 11
\begin{figure}
\epsscale{1.} 
\plotone{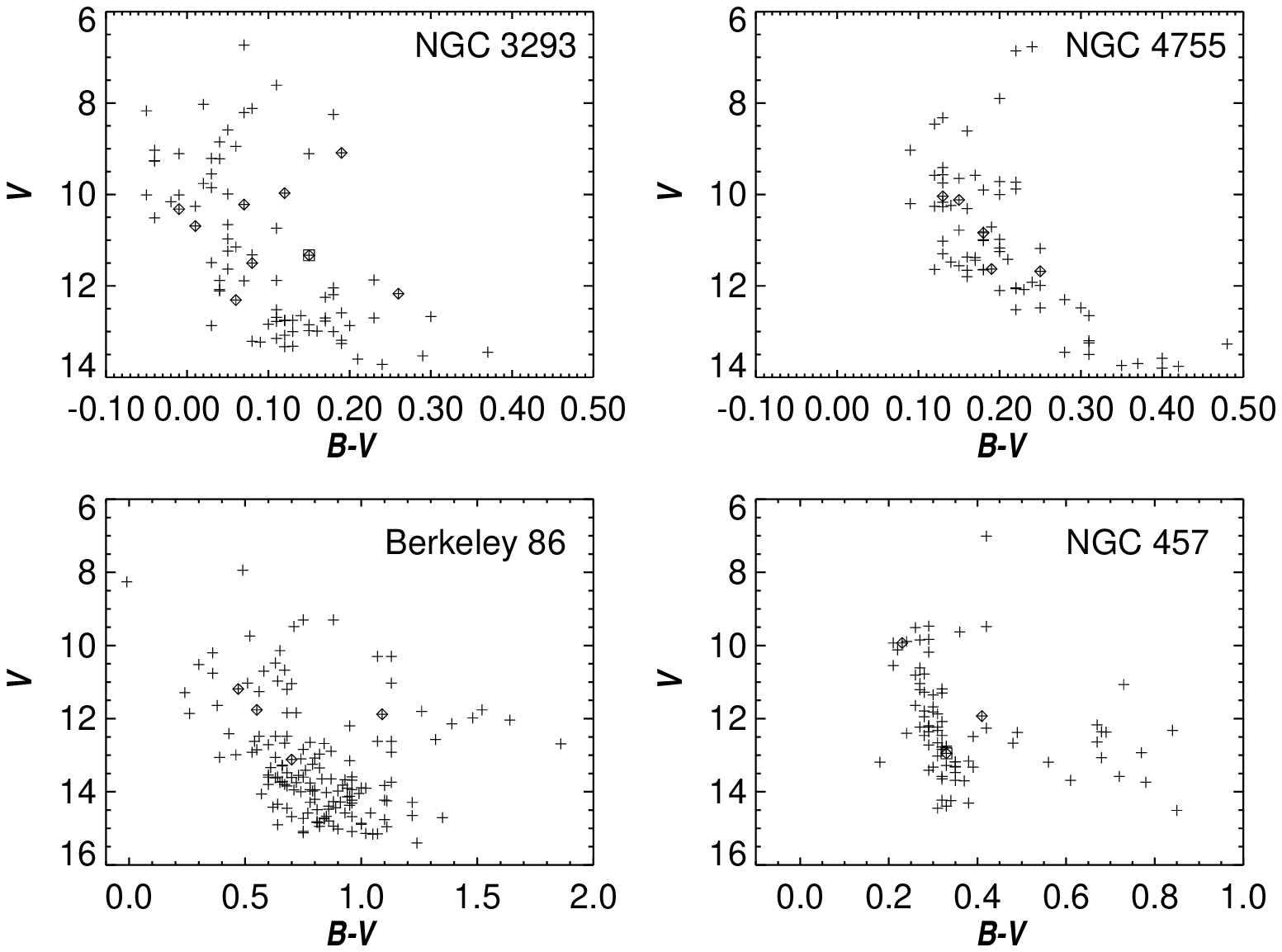}
\caption{
Color-magnitude diagrams for
NGC 3293 (from photometry by \citealt{tur80}),
Berkeley 86 (photometry by \citealt*{for92} and \citealt*{mas95}), NGC 4755 (photometry by
\citealt{per76} and \citealt{dac84}),
and NGC 457 (photometry by \citealt{pes59} and \citealt{hoa61}).
The crosses enclosed in diamonds mark the fast rotators ($V \sin i > 260$
km~s$^{-1}$).  The only two candidate binary stars identified among 
the fast rotators are NGC~3293 \#83 and NGC 457 \#43, which are marked 
by enclosing squares.  
}
\label{fig11}
\end{figure}

%%%%%%%%%%%%%%%%%%%%%%%%%%%%%%%%%%%%%%%%%%%%%%%%%%%%%%%%%%%%%%%

\end{document}